\documentstyle[12pt,epsf]{article}
\newcommand{\Section}[1]%
{\section{#1}\setcounter{equation}{0}%
\setcounter{theorem}{0}}

\begin{document}
\newpage\thispagestyle{empty}
\topskip 2.5cm
\vskip 2cm
\begin{center}
{\Large\bf
Low-lying excitations
around a single vortex\\
in a d-wave superconductor\\}
\bigskip\bigskip
{\large\bf Y. Morita$^{1}$\\}
\bigskip
{\large\bf M. Kohmoto$^{2}$$^{\dagger}$\\}
\bigskip
{\large\bf K. Maki$^{3}$\\}
\end{center}
\vfil
A full quantum-mechanical treatment of
the Bogoliubov-de Gennes equation
for a single vortex in a d-wave superconductor
is presented.
First,
we find
low-energy states
extended in four diagonal directions,
which have no counterpart
in a vortex of s-wave superconductors.
The four-fold symmetry is
due to '{\it quantum effect}',
which is enhanced
when $p_{F}\xi $ is small.
Second, for $p_{F}\xi \sim 1$,
a peak with a {\it large energy gap}
$E_{0}\sim \Delta $
is found in the density of states,
which is due to
the formation of
the lowest bound states.
\par\noindent
\medskip
\hrule
\bigskip
\noindent
\medskip

\noindent
$^{1}$
{\small Department of Applied Physics, University of Tokyo,
7-3-1 Hongo Bunkyo-ku, Tokyo 113, Japan
}
$^{2}$
{\small Institute for Solid State Physics,
University of Tokyo
7-22-1 Roppongi Minato-ku, Tokyo 106, Japan
}
$^{3}$
{\small Department of Physics and Astronomy,
University of Southern Calfornia Los Angeles,
Cal. 90089-0484, USA
}
$^{\dagger}$
{\small Author to whom
all correspondence should be addressed.}
{\small\tt E-mail:kohmoto@issp.u-tokyo.ac.jp}\\
\noindent
\newpage
After a few years of controversy,
d-wave nature of the high-T$_{c}$ superconductors
is now well estabilished \cite{exp1,exp2},
although
superconductivity in the electron-doped
Nd$_{2-x}$Ce$_{x}$CuO$_{4}$ appears to be of s-wave \cite{exp3}.
Therefore
it is important
to understand
the nature of vortex states
in a d-wave superconductor \cite{exp2,maki5}.
An earlier analysis of the vortex state
based on the Gor'kov equation
shows
that
a square lattice of vortices
tilted by ${\pi}/4$ from the a-axis
is the most stable
except
in the immediate vicinty of $T=T_{c}$
or in a weak magnetic field \cite{maki2}.
Such a square lattice of vortices, though distorted,
has been seen
by
a small angle neutron scattering \cite{neutron}
and
a scanning tunneling microscopy (STM) \cite{geneve}
in YBa$_{2}$Cu$_{3}$O$_{7-{\delta}}$ (YBCO)
at low temperatures ($T<10$K).
We believe that
this distortion of the vortex lattice
is due to
the orthorhomibicity of the YBCO,
although there are alternative interpretations
based on the (d+s) admixture \cite{sd1,sd2}.
One of the most remarkable results
in the STM experiment is that
the vortex appears to have a circular symmetry
as in an s-wave superconductor.
It is in sharp contrast to earlier results
obtained
within the Eilenberger theory
(a semi-classical theory of a superconductor)
\cite{eilen},
where
a clear four-fold symmetry was obtained
in the local density of states
\cite{maki1,maki4,kyoto}.
Further,
at the center of the vortex,
a peak with a {\it large energy gap}
$E_{0}\sim \Delta $
was found in the local density of states,
where $\Delta$ is
the superconducting order parameter.
Then the most natural interpretation
is that
this corresponds to the lowest bound state
for a vortex in a d-wave superconductor
analogous to the one
predicted by Caroli, de Gennes and Matricon
\cite{degennes}.

In the previous study \cite{morita},
in order to understand
the results from the STM experiment,
we have solved
the Bogoliubov-de Gennes (B-dG) equation
for a d-wave superconductor
and
obtained
quasi-particle spectra
around a single vortex.
In the temperature region
where the Ginzburg-Landau (GL) theory is valid,
we found that
the local density of states
exhibits a circular symmetry and
a peak with a {\it large energy gap}
$E_{0}\sim \Delta$
is found
in the local density of states
at the center of the vortex,
which is consistent with the STM experiment.
In ref. \cite{morita},
it is crucial to set $p_{F}\xi \sim 1$ for YBCO,
where
$p_{F}$ and $\xi$ are
the Fermi momentum and the coherence length
respectively.
The value of $p_{F}\xi$ is obtained
by an approximate formula
for the lowest bound state
$E_{0}={\Delta}/(\pi p_{F}\xi)$ \cite{maki4,sbdg1}.
This is also consistent with the chemical potential of YBCO
deduced from the analysis of the spin gap
seen
in an inelastic neutron scattering experiment
from monocrystals of YBCO \cite{RM,maki6}.
In the above analysis, however,
we neglected the {\it noncommutability
between $\hat k$ and $\bf x$
('quantum effect')}{\cite{bruder}},
where
$\hat k$ and $\bf x$
are
the quasi-particle
momentum
and
coordinate
respectively,
and
the local density of states
has a {\it perfect} circular symmetry
except when the mixing of an s-wave component occurs \cite{ren1}.
The correction
is $O(1/p_{F}{\xi})$
and
irrelevant
at least in the study of
systems with a long cohernce length
{\it e.g.}
the superconducting phases
of the heavy-ferimon systems
($p_{F}{\xi}\sim 10$)
and
the ${}^{3}$He superfluidity
($p_{F}{\xi}\sim 100$),
but may have a serious influence
in the study of
the high-$\rm T_{c}$ superconductors,
where $p_{F}{\xi}\sim 1$ as is discussed above.

In this paper,
a full quantum-mechanical treatment
of the B-dG equation for a d-wave superconductor
is reported,
where
the
{\it 'quantum effect'}
is taken into account.
As shown below,
the four-fold symmetry appears
in the local density of states.
Similar four-fold symmetry was obtained
in the previous studies \cite{maki1,maki4,kyoto}.
But it should be noted that
the four-fold symmetry discussed here
has totally different origin
from that obtained in the earlier studies.
The B-dG equation for a d-wave superconductor
is given by
\begin{eqnarray}
&&
\Bigl \{
-{\frac{1}{2m}}
({\nabla}-ie{\bf {A}}({\bf x}))^{2}-{\mu }
\Bigr \}
u_{n}({\bf x})
\nonumber
\\
&&-
{\{}
{\partial}_{x}({\Delta}({\bf x}){\partial}_{x})-
{\partial}_{y}({\Delta}({\bf x}){\partial}_{y})
{\}}
v_{n}({\bf x})
=
{\epsilon }_{n}
u_{n}({\bf x}),
\nonumber
\\
&&
-
\Bigl \{
-{\frac{1}{2m}}
({\nabla}+ie{\bf {A}}({\bf x}))^{2}-{\mu }
\Bigr \}
v_{n}({\bf x})
\nonumber
\\
&&-
{\{}
{\partial}_{x}({\Delta}({\bf x})^{*}{\partial}_{x})-
{\partial}_{y}({\Delta}({\bf x})^{*}{\partial}_{y})
{\}}
u_{n}({\bf x})
=
{\epsilon }_{n}
v_{n}({\bf x}),
\nonumber
\\
&&
\nonumber
\label{bdg1}
\end{eqnarray}
where
$u_{n}({\bf x})$
and $v_{n}({\bf x})$
are
the quasi-particle amplitudes,
${\Delta}({\bf x})$
is the pair potential,
${\bf {A}}({\bf x})$ is the vector potential
which is neglected assuming $H{\ll}H_{c2}$,
and
$\mu$ is the chemical potential
which is identified with the Fermi energy.
The parameters are set as
$2m{\xi}^{2}\Delta =2.82$
and
$R/\xi=30$,
where
$\xi$ is the coherence length
and
$R$ is radius of the system
and the pair potential
${\Delta}({\bf x})$
is given by
\begin{equation}
{\Delta}({\bf x})
={\Delta}\tanh (r/{\xi })e^{i{\phi}},
\nonumber
\label{test1}
\nonumber
\end{equation}
where
$r$ and $\phi$
are defined by
${\bf x}=(r\cos(\phi),r\sin(\phi))$.
This form of pair potential
is obtained
by the Ginzburg-Landau (GL) theory
for a d-wave superconductor
\cite{maki4,ren1}
and applicable at not too low temparatures
(an estimate of the temperature region
where the GL theory is valid is
$[0.5T_{c},T_{c}]$ for
an s-wave superconductor,
when one consider the quasi-particle spectra
around a single vortex).

In order to solve the B-dG equation
numerically,
it is convenient to expand
the quasi-particle amplitudes
$u_{n}({\bf x})$
and
$v_{n}({\bf x})$
as
\begin{eqnarray}
&&
u_{n}({\bf x})
=
{\sum_{l=-\infty}^{\infty}}
{\sum_{j=1}^{\infty}}
u_{n,l,j}{\psi}_{j,|l|}(r)\exp(il{\phi}),
\nonumber
\\
&&
v_{n}(\hat{k})
=
{\sum_{l=-\infty}^{\infty}}
{\sum_{j=1}^{\infty}}
v_{n,l,j}{\psi}_{j,|l-1|}(r)\exp(i(l-1){\phi}).
\label{mode}
\nonumber
\end{eqnarray}
Here
${\psi}_{i,\nu}(x)=
\frac{1}{{\sqrt{2\pi}}RJ_{\nu+1}(\alpha _{i,\nu})}
J_{\nu}(\alpha _{i,\nu}x/R)$
($J_{\nu}(x)$
is the Bessel function),
$\alpha _{j,\nu}$ is the $j$-th positive zero point
of $J_{\nu}(x)$
and $R$ is the radius of the system.
In the previous study \cite{morita},
in which
the {\it 'quantum effect'}
is neglected,
the B-dG equation decouples
for $u_{n,l,j}$'s and $v_{n,l,j}$'s
with different $l$.
On the other hand,
when the {\it 'quantum effect'}
is taken into account,
$u_{n,l,j}$'s and $v_{n,l,j}$'s
with different $l$
couple,
which plays an important role
when $p_{F}\xi$ is small.
Because of the coupling,
the number of basis
we can use for numerical diagonalization
is small compared to the previous study \cite{morita}.
However, it is sufficient
for the understanding of
the qualitative aspects.

At first, consider
the density of states
$\sum _{i}\delta(E-E_{i})$
as a function of $E/{\Delta}$,
where $p_{F}\xi =1.33$.
As is seen in Fig. 1,
there is a peak with a {\it large energy gap}
$E_{0}\sim \Delta $.
This corresponds to the lowest bound state.
The peak has a width
due to the internal degree of freedom
in the $\hat k$ space.
These are consistent
with the previous results \cite{morita} qualitatively,
where the {\it 'quantum effect'} is neglected.

Next consider a local density of states in a superconductor,
which is the quantity of interest
for comparison to STM experiments and given by,
\begin{eqnarray}
N(E,{\bf x})
&&=
\sum_{n}
[\
|u_{n}({\bf x})|^{2}
\ \delta (E-{\epsilon _{n}})
\nonumber
\\
&&+
|v_{n}({\bf x})|^{2}
\ \delta (E+{\epsilon _{n}})
].
\label
{ldos}
\nonumber
\end{eqnarray}
In Fig. 2 and 3,
$\int _{0}^{0.35\Delta} dE\ N(E,r,\phi)$
is plotted and
they
show a clear four-fold symmetry
in the local density of states.
When $p_{F}\xi$ is changed from 1.33 to 2.00,
the four-fold symmetry is suppressed (see Fig. 3)
and
the local density of states becomes circular.
This supports the idea that
the four-fold symmetry is due to the {\it 'quantum effect'}.
We stress that
these low-energy states
extended in four diagonal directions
are paticular to d-wave superconductivity.
Therefore
1)
these states should
give rise to
the zero-energy density of states
proportional to $\sqrt{B}$ as
discussed by Volovik and others \cite{vol,maki5},
2)
they are the most likely
the origin of the large flux flow resitivity
in YBCO observed recently by Doettinger et al. \cite{flow}
and,
3)
when a square lattice of vortices
tilted by $\pi/4$ from the a-axis is formed,
the quasi-paricle
can move
from one vortex to the other through
these low-energy states extended in four diagonal directions,
which should give rise to a cohesive energy
guaranteeing the stability of the square lattice.
The clarification of the quasi-particle spectrum
in a vortex lattice
and, in paticular, the tilted square lattice
is of immediate interest.

In conclusion,
we have investigated
the Bogoliubov-de Gennes equation for a d-wave superconductor,
where
the {\it noncommutability between $\hat k$ and $\bf x$
('quantum effect')}
is taken into account.
We found a peak with a {\it large energy gap}
$E_{0}\sim \Delta $
in the density of states,
which is consistent with the previous results.
We found low-energy states extended
in four diagonal directions,
which is due to the {\it 'quantum effect'}.
The low-energy states have no counterpart
in a vortex of s-wave superconductors.
It is natural to consider
that
these low-energy states
cause directional attractive forces between vortices.
It is possible that,
due to the directional attractive force,
a square lattice of vortices becomes stable
in some parameter region.
Another scenario
for a square lattice of vortices
is proposed in ref. \cite{maki2},
where
the higher-order correction
in the Ginzburg-Landau theory
\cite{glcor}
plays an essential role.
In this paper,
we do not consider
the effect of the higher-order correction.
The higher-order correction
causes the four-fold symmetry
in the pair potential.
We consider that, in low temperatures,
it is needed
to take into account
both
the {\it 'quantum effect'}
and
the higher-order correction in the GL theory
self-consistently,
and
more detailed study
is left as a future problem.

This work is in part supported
by National Science Foundation
under grant number DMR95-31720,
ISI Foundation at Torino, Italy
and
timely short term research fellowship
of Japan Society of Promotion of Science.

\newpage
\noindent
{\large \bf Figure Captions}

\begin{description}

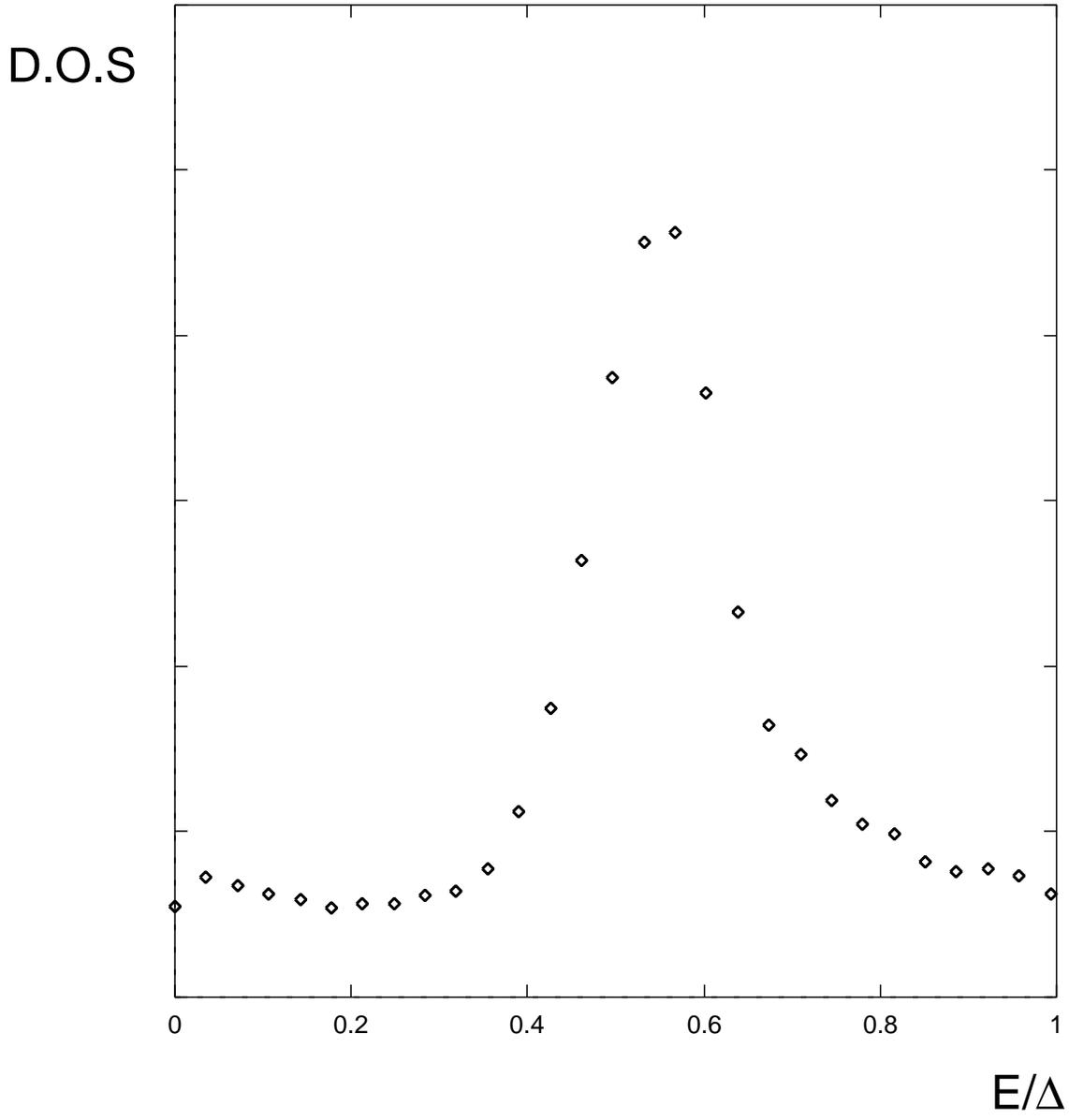
\item{Fig.~1:}
$\sum _{i}\delta(E-E_{i})$
as a function of $E/{\Delta}$,
where $k_{F}\xi =1.33$.

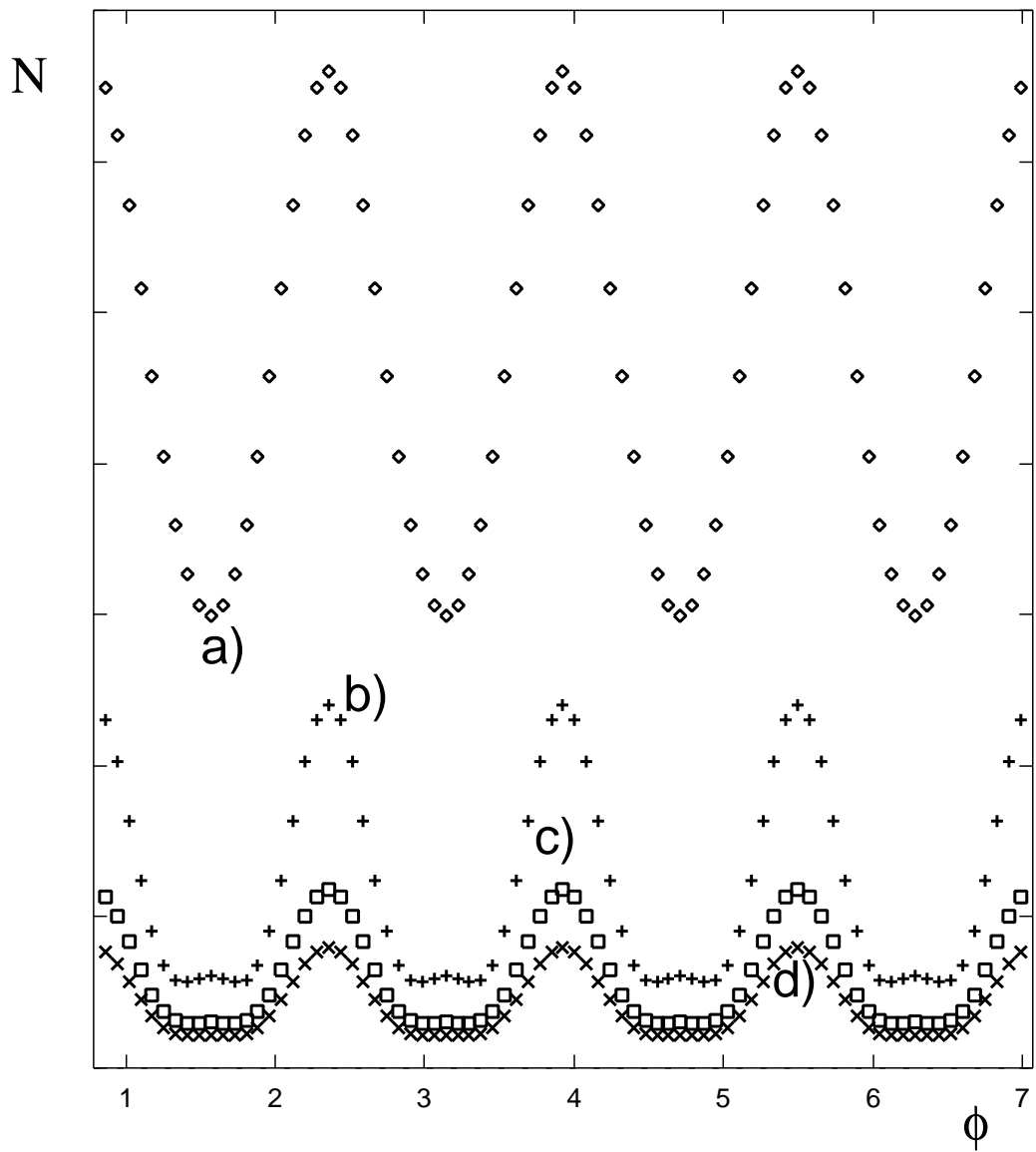
\item{Fig.~2:}
$\int _{0}^{0.35\Delta} dE\ N(E,r,\phi)$,
where $k_{F}\xi =1.33$
for
a) $r/{\xi}=3.0$,
b) $r/{\xi}=9.0$,
c) $r/{\xi}=15.0$
and
d) $r/{\xi}=27.0$.

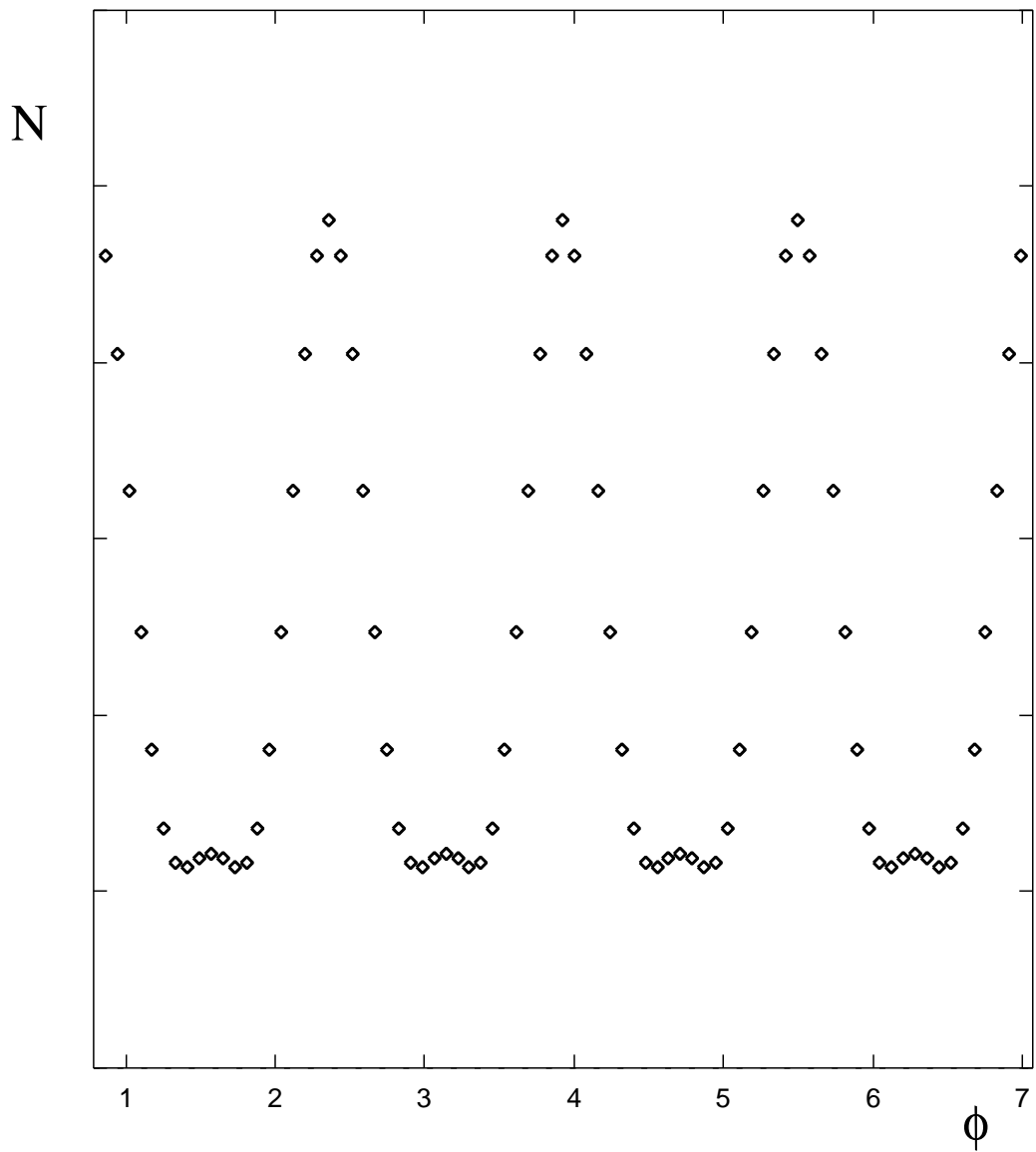
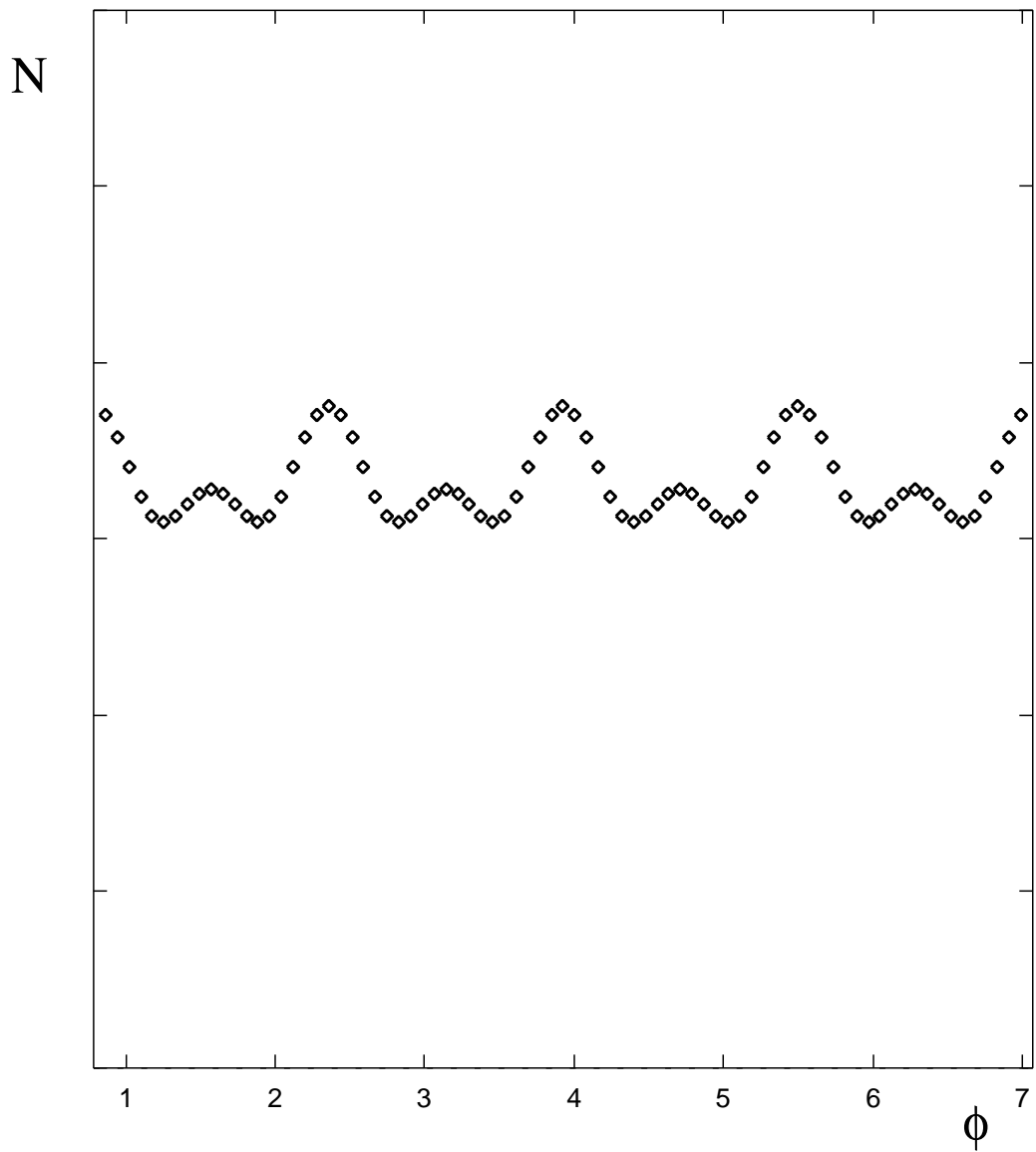
\item{Fig.~3:}
$\int _{0}^{0.35\Delta} dE\ N(E,r,\phi)$,
where
a)
$k_{F}\xi =1.33$ and $r/{\xi}=9.0$
and
b)
$k_{F}\xi =2.0$ and $r/{\xi}=9.0$.

\end{description}

\end{document}